\RequirePackage{ifpdf}
\ifpdf 
\documentclass[pdftex]{sigma}
\else
\documentclass{sigma}
\fi

\begin{document}
\allowdisplaybreaks

\def\dx#1{{\partial \over \partial#1}}
\def\dh#1{\mathop {#1}\limits_{h}}
\def\dhp#1{\mathop {#1}\limits_{+h}}

\def\dhm#1{ \mathop{#1}\limits_{-h}}
\def\dphh#1{ \mathop{#1}\limits_{h \bar h}}
\def\da#1{ \mathop{#1}\limits_{+\tau}}
\def\db#1{ \mathop{#1}\limits_{-\tau}}
\def\dc#1{ \mathop{#1}\limits_{\pm \tau}}
\def\dd#1{ \mathop{#1}\limits_{+h}}
\def\df#1{ \mathop{#1}\limits_{-h}}
\def\dpm#1{ \mathop{#1}\limits_{\pm h}}
\def\dg#1{ \mathop{#1}\limits_{\pm h}}
\def\dh#1{ \mathop{#1}\limits_ h}

\def\sso#1{\ensuremath{\mathfrak{#1}}}

\newcommand{\ddt}{\partial \over \partial t}
\newcommand{\ddx}{\partial \over \partial x}
\newcommand{\ddy}{\partial \over \partial y}
\newcommand{\ddyy}{\partial \over \partial y'}

\renewcommand{\PaperNumber}{065}

\FirstPageHeading

\ShortArticleName{On the Linearization of Second-Order
Dif\/ferential and Dif\/ference Equations}

\ArticleName{On the Linearization of Second-Order Dif\/ferential\\
and Dif\/ference Equations}

\Author{Vladimir DORODNITSYN} 
\AuthorNameForHeading{V. Dorodnitsyn}

\Address{Keldysh Institute of Applied Mathematics of Russian
Academy of Science,\\ 4 Miusskaya Sq., Moscow, 125047 Russia}

\Email{\href{mailto:dorod@spp.keldysh.ru}{dorod@spp.keldysh.ru}}

\ArticleDates{Received November 28, 2005, in f\/inal form July 13,
2006; Published online August 16, 2006}

\Abstract{This article complements  recent results of the papers
[{\it J.~Math.~Phys.} 41 (2000), 480; 45 (2004), 336]
on the symmetry classif\/ication of second-order ordinary
dif\/ference equations and meshes, as well as the   Lagrangian
formalism and
 Noether-type integration technique.
    It turned out that there exist nonlinear
 superposition principles for  solutions of special  second-order ordinary
 dif\/ference equations which possess Lie group symmetries.
 This superposition springs from the linearization of second-order ordinary
 dif\/ference equations by means of non-point transformations which act simultaneously
 on equations and meshes. These transformations become some sort of contact
 transformations in the continuous limit.}


 \Classification{34C14; 34C20; 39A05; 65L12; 70H33}

\section{Introduction}

\looseness=1 The recent article \cite{1} was devoted to symmetry
classif\/ication of second-order ordinary dif\/ference equations.
This was based on the paper by S.~Lie \cite{3}, in which he has
provided a symmetry classif\/ication of second-order
dif\/ferential equations (ODEs).
  A special class of the specified invariant
 second-order dif\/ference equations \cite{1} converges to
Lie's invariant second-order ordinary dif\/ferential equations in
the continuous limit. The Lagrangian formalism for Lie's list of
those second-order ordinary dif\/ference equations associated with
special invariant meshes (lattices) was developed in recent
paper~\cite{2}. From the Noether theorem \cite{2} it is known that
invariance of a variational functional with respect to an
$r$-parameter group leads to existence of $r$ conservation laws
for the corresponding Euler's equation. It was shown in \cite{2}
(see also \cite{5,6} for details) that there exists a similar
(although more complicated) construction for dif\/ference models.
This discrete analog of the Noether theorem can be applied to
analytic integration of dif\/ference equations. Namely,
two-dimensional symmetries of the Lagrangian for a second-order
dif\/ference equation provide two f\/irst dif\/ference integrals
of the two-point type. This yields a general solution of the
dif\/ference system by purely algebraic manipulations. The last
point is  important for dif\/ference equations since there are  no
analytic integration techniques for constructing solutions.

The problem of  linearization of a given equation (or a system of
equations) is of permanent interest in mathematics. For example,
in the paper \cite{7} necessary and suf\/f\/icient conditions,
under which a given nonlinear system of PDEs with $n \geq 2$
independent variables and $m \geq 1$ dependent variables can be
transformed to a linear system of PDEs, were developed. It was
shown that in all the cases of such mappings, an {\it
infinite-parameter} Lie group of point or contact transformations
admitted by the original nonlinear system  is needed.

For ODEs the situation is dif\/ferent. S.~Lie proved that symmetry
algebras admitted by second-order ODEs are {\it finite},  and the
maximum symmetry algebra is 8-dimensional. This maximum symmetry
algebra is admitted by a linear equation. S.Lie has developed
necessary and suf\/f\/icient conditions under which a given
nonlinear second-order ODE can be transformed into a linear one
\cite{8} (the discussion of this question can be found, for
example, in \cite[p.~38]{9}. One of these (equivalent to each
other) conditions is existence of two linearly connected  symmetry
operators. In this case there exists a change of variables that
transforms a given nonlinear equation into a linear one.
Generalization of this idea to
 ordinary dif\/ference equations was developed in \cite{10} where, for
 some class of ordinary dif\/ference equations on f\/ixed uniform
 lattice, point transformations which linearize
 corresponding nonlinear dif\/ference equations have been found.
 This class of ordinary dif\/ference
 equations was restricted by those equations that possess  point autonomous
  symmetries, which do not change the  independent  variable.
  Dif\/ference equations are assumed to be def\/ined on a f\/ixed regular
 lattice with unit spacing. This approach was generalized in \cite{11} to
 non-autonomous symmetries, while the independent variable was still f\/ixed.
The discrete analog of Lie's theorem \cite{8}, providing the
necessary and suf\/f\/icient conditions, under which a given
nonlinear second-order dif\/ference equation and a mesh can be
transformed into a linear one, was developed in \cite[Theorem
1]{1}.
 It is important to notice that in all the cases
the corresponding {\it point transformations} change equations
within one class of equivalent dif\/ference equations, which are
invariant under the actions of {\it isomorphic} symmetry groups.
Point transformations of such type cannot connect  nonlinear
equations  from  Lie's list with a linear one. Being based on
listing of groups, the symmetry classif\/ication of second-order
dif\/ferential \cite{3} and dif\/ference equations \cite{1}
singled out classes of equations, which cannot be connected  by
means of point transformations (of independent and dependent
variables) and, in particular,  cannot be linearized by a point
transformation in principle. In the classif\/ication \cite{3}
S.~Lie singled second-order ODEs up to {\it arbitrary point
transformations}. For example, a linear equation is the simplest
representative equation from the class of equivalent equations
which possess an 8-dimensional symmetry group.

Thus, in this article we are dealing with ``linearization of
not-linearizable equations'' in the above mentioned sense. We
establish {\it non-point} transformations connecting some
nonlinear invariant second-order dif\/ferential equations from
Lie's list with linear equations. Such transformations have
discrete counterparts,  and so the invariant dif\/ference schemes
can be linearized together with dif\/ference meshes.

We will use the following notation. Let $x$ be the independent
variable and $y$ the dependent variable. To consider a
second-order dif\/ference equation and a lattice in $x$-direction
we need 3-points of a dif\/ference mesh (lattice):
$(x,x_{-},x_{+})$ and corresponding values of the dependent
variable in three neighboring points: $(y,y_{-},y_{+})$, where
``$+$'' marks a point shifted to the right, and ``$-$'' marks a
point shifted to the left. We call this set of points (a subspace)
$(x,x_{-},x_{+},y,y_{-},y_{+})$ a {\it dif\/ference stencil}. It
is supposed that the stencil and the corresponding dif\/ference
equation can be shifted to any other point of the lattice.

The discrete model for second order ODE can be presented in terms
of two dif\/ference equations:
\begin{gather}\label{eq1}
 F ( x,x_{-},x_{+},y,y_{-},y_{+} ) = 0,\qquad
 \Omega (x,x_{-},x_{+},y,y_{-},y_{+}) = 0.
\end{gather}

We will also use the following notation for the right and for the
left dif\/ference derivatives of the f\/irst order:
\[
 y_{x} = { y_{+} -y \over h_{+} }, \qquad
 y_{\bar{x}} = { y - y_{-}  \over h_{-} }.
\]
where $h_{+}$, $h_{-}$ are the space steps of the mesh in the
$x$-direction.

To distinguish between continuous derivatives and dif\/ference
ones we denote the f\/irst by $y', y'',\ldots$. The f\/irst
equation of the system \eqref{eq1} represents a second-order
dif\/ference equation. In the continuous limit it should become a
second-order ordinary dif\/ferential equation on some point of a
f\/inite-dimensional space of continuous variables. The second
equation yields a lattice (mesh) on which the f\/irst equation is
considered. Notice that the mesh equation does not need to be a
second-order dif\/ferential equation in the continuous limit. For
example, one can consider a uniform lattice equation $h_{+}=
h_{-}$ (where $h_{+} = x_{+}  - x$, $h_{-} = x - x_{-}$) which
``disappears'' in the continuous limit.

The group transformations considered in the present approach are
of the same type as for ODEs. They are generated by a Lie algebra
of vector f\/ields of the form
\begin{gather*} 
 X =   \xi (x,y) { \ddx } + \eta (x,y) {\ddy }.
\end{gather*}

The corresponding transformations are purely point ones, since the
coef\/f\/icients $ \xi$ and $\eta$ depend  on $(x,y)$ only, and do
not depend on the shifted points  $(x_{+}, y_{+})$ or
$(x_{-},y_{-})$.

In the classif\/ication of second-order ODEs \cite{3} S.~Lie
identified
 a linear equation, which possess 8-dimensional symmetry group; four
 simplest representative nonlinear equations which possess a 3-dimensional
 symmetry group; 2 classes of ODEs with a 2-dimensional
 symmetry group and one class of ODEs with a one-dimensional
 symmetry group.

 In the following sections we consider three examples of ODEs that
possess three symmetries. In  parallel we will consider the
corresponding dif\/ference models as developed in \cite{2}. Also
we consider {\it examples} from the Lie classes of nonlinear ODEs
with a two-dimensional  Lie algebra. The chosen equations
 are important from the physical point of view.

 We will write down explicitly the linearizing tangent transformations
 for underlining
second-order nonlinear ODEs. We will also write down explicitly
the linearizing two-points transformations for corresponding
dif\/ference equations and meshes. Notice that these non-point
transformations change the admitted symmetries of the underlying
equations as the corresponding linear equations possess the
8-dimensional symmetry. In fact, these transformations are not
explicitly connected with admitted symmetries of ODEs. This aspect
will be discussed later.

Notice that all dif\/ference and dif\/ferential equations
considered here are integrable (at least in quadratures) as far as
they possess a symmetry algebra of dimension not less than~2.
Therefore, the computational aspects of the invariant dif\/ference
schemes (as well as stability, convergence of developed schemes,
etc.) will not discussed here.

In the present paper we are following the set of invariant ODEs
for which an appropriate invariant dif\/ference equations and
meshes were developed in \cite{1,2}.

It seems  that the main result presented below is the {\it
explicit non-point transformations}, which linearize second-order
ODEs from the Lie's list and appropriate invariant dif\/ference
equations, which possess symmetries of the dimension less than~8.
A discussion of the origin of such transformations will be
presented at the end of  Example~1 and in Concluding remarks.

\section{Example 1: ODE with a 3-dimensional symmetry group}

We start with the example of a second-order nonlinear ODE from the
Lie's list of equations, which possesses three symmetries. It has
two f\/irst integrals and, therefore, can be integrated by means
of algebraic manipulations with dif\/ference integrals.

The group given by the operators
\begin{gather} \label{op32}
X_{1}= {\dx x} , \qquad  X_{2}= 2 x {\dx x} + y {\dx y} , \qquad
X_{3}= x^{2} {\ddx} + xy {\ddy},
\end{gather}
corresponds to the invariant dif\/ferential equation from the
Lie's list,
\begin{gather} \label{eq3.14}
y'' =  y^{-3}.
\end{gather}

This equation can be obtained from a variational functional with
the Lagrangian function $ L = {y'}^2   - { 1 \over  y ^2 }$ which
admits all three operators \eqref{op32} as variational symmetries.

 The Noether theorem \cite{4} yields the following f\/irst integrals:
\begin{gather} \label{lagh32}
J_{1} =  {y'}^2   + { 1 \over  y ^2 } = A_0  , \qquad J_{2} =   {x
\over y^2} - (y - y'x) y' = B_0  .
\end{gather}

By means of the f\/irst integrals we write the general solution
$y(x)$ as
\begin{gather} \label{sol3aa}
A_0 y^{2} = ( A_0 x - B_0 )^{2} + 1.
\end{gather}

Now we reproduce the procedure of the linearization for the
equation \eqref{eq3.14}, followed by explanatory comments.

 It  turns
out that  equation \eqref{eq3.14} can be linearized by the
non-point transformation
\begin{gather} \label{iso1}
t= x\left({y'}^2   + { 1 \over  y ^2 }\right)+ \arctan(yy')-yy',
\qquad y(x) = \frac{1}{u(t)}, \qquad y'= -u',
\end{gather}
which transforms  equation \eqref{eq3.14} into the following
linear equation
\begin{gather} \label{iso2}
u''+u =0,
\end{gather}
which possesses a linear superposition for its solutions.

Thus, the developed formulas implicitly reproduce {\it the
 nonlinear superposition principle} for equation \eqref{eq3.14}.
The transformation \eqref{iso1} changes the f\/irst integrals
\eqref{lagh32} into the following ones with the same constants
$A_0, B_0$:
\begin{gather*} 
{u'}^2 + u^2 =A_0, \qquad  t+\arctan \left(\frac{u'}{u}\right)
=B_0.
\end{gather*}
The latter yields the general solution of the linear equation
\eqref{iso2}
\begin{gather*} 
 u(t)= \sqrt{A_0} \cos (B_0 - t)= \tilde{A} \sin t + \tilde{B} \cos t.
\end{gather*}

Notice that \eqref{iso1} connects the general solution
\eqref{sol3aa} with (10), which provides an explicit procedure for
the superposition of solutions \eqref{sol3aa}. Indeed, let us take
two solutions among the family~\eqref{sol3aa}, which correspond to
the two sets of constants: $ y_1 = Y_1 (x,A_1,B_1)$ and  $ y_2 =
Y_2 (x,A_2,B_2)$. Then the transformation \eqref{iso1} produces
two  solutions of the linear equation \eqref{iso2},
\begin{gather*} 
 u_{1}(t)= \sqrt{A_1} \cos (B_1 - t), \qquad u_{2}(t)= \sqrt{A_2} \cos (B_2 - t).
\end{gather*}

The sum of these solutions
\begin{gather*} 
 u_{}(t)=u_1 + u_2= \sqrt{A^{*}} \cos (B^{*} - t),\\
 {A^{*}}=A_1 + A_2 +
 2\sqrt{A_1 A_2} \cos (B_1 - B_2),\qquad
 B^{*} = \arctan {\frac{ \sqrt{A_1} \sin B_1+
 \sqrt{A_2} \sin B_2}{\sqrt{A_1} \cos B_1+
 \sqrt{A_2} \cos B_2}},\nonumber
\end{gather*}
 yields a new solution of  equation \eqref{eq3.14} with two constants $
 A^{*}$, $B^{*}$. The developed formulas reproduce  the
 nonlinear superposition principle within the family of solutions
 \eqref{sol3aa}. For example, the solutions $ {y_{1}} =\sqrt{x^2 +1}$ and
$ {y_{2}} =\sqrt{4x^2 +{1\over{4}}}$ produce the new one  $
{y_{3}} =\sqrt{7x^2 +{1\over{7}}}$.

Now we consider the discrete case. In the paper \cite{2} a
dif\/ference scheme which preserves symmetries and the Lagrangian
structure of the initial equation \eqref{eq3.14} was developed.

The starting point was an entire set of f\/inite-dif\/ference
invariants
\[
I_1= {h_{+} \over y y_{+} }, \qquad I_2 ={h_{-} \over y y_{-} },
\qquad  I_3 =\frac{y^2 y_{-}}{h_{-}}\left(\frac{y_+ -y   }{ h_{+}
}-\frac{y -y_{-} }{ h_{-} }\right).
\]
in a subspace of dif\/ference variables (dif\/ference stencil)
$x$, $h_+$, $h_-$, $y$, $y_{+}$, $y_{-}.$

Any dif\/ference equation which can be represented by means of the
above invariants will be invariant. Thus, in such a way one can
generate an invariant mesh. We will use the following relation
between dif\/ference invariants
\begin{gather} \label{gridss2}
 {h_{+} \over y y_{+} } ={h_{-} \over y y_{-} }
\end{gather}
as the invariant mesh (see \cite{2}).

 It is easy to see that the mesh equation has a f\/irst integral
\begin{gather*} 
 {h_{+} \over y y_{+} } = \varepsilon,\qquad \varepsilon = {\rm const}.
\end{gather*}

As an invariant representation of the equation \eqref{eq3.14} we
consider the following:
\begin{gather}\label{eq15}
\frac{1}{h_{-}}\left(\frac{y_+ -y   }{ h_{+} }-\frac{y -y_{-}  }{
h_{-} }\right) ={1 \over y^2 y_{-} }.
\end{gather}
The last equation on the mesh \eqref{gridss2} can be rewritten as
the mapping
\begin{gather*}
y_{+}y (2- \varepsilon^{2}) =y(y_{+} + y_{-}).
\end{gather*}

Notice that dif\/ference model \eqref{gridss2}, \eqref{eq15} of
equation \eqref{eq3.14} can be developed from the discrete
Lagrangian function $ {\cal L} = y_{x}^2 - { 1 \over y y^{+}}  $
which
 gives  f\/irst integrals by means of a f\/inite-dif\/ference analog of the
 Noether theorem \cite{5,6}:
\begin{gather} \label{int2}
I_{1} =  { y_x}^2 + \frac{1}{y y^{+} } = A , \qquad I_{2} =
\frac{x + x^{+} }{ 2 y y^{+}  +  { y_x} ( x^{+}  y_x - y^+  )} =
B.
\end{gather}
Eliminating $y_x$, $x_+$ and $y_+$ from the  f\/irst dif\/ference
integrals, we obtain the general solution
\begin{gather} \label{sol22}
Ay^2 = ( Ax - B ) ^2 + 1 - { \varepsilon ^2 \over 4 }  ,
\end{gather}
which agrees with the continuous limit  up to order $ \varepsilon
^2 $.

To obtain the dif\/ference analog of the non-point transformation
\eqref{iso1} we rewrite the latter as relations between the
dif\/ferentials:
\begin{gather*} 
 dx= \frac{dt}{u^2}, \qquad  y(x) =\frac{1}{u(t)} .
\end{gather*}
Then, as a dif\/ference analog of the above transformation, we
will use non-point transformations for discrete variables as
follows
\begin{gather} \label{sal8}
 y =\frac{1}{u}, \qquad y^+ =\frac{1}{u^+},\qquad y^- =\frac{1}{u^-},
\qquad \tau^{+}=\frac{h^+}{yy^+},\qquad
\tau^{-}=\frac{h^-}{yy^-},\qquad  y_x=-u_t.
\end{gather}
where $\tau^{+}$ and $\tau^{-} $ are the mesh steps in the new
coordinate system. Invariant mesh \eqref{gridss2} is transformed
into the following regular mesh in the new coordinate system:
\begin{gather*} 
\tau^{+}=\tau^{-}=\tau_{0}=\varepsilon,
\end{gather*}
and the nonlinear equation \eqref{eq15} becomes linear:
\begin{gather*} 
\frac{u^+ -2u + u^-}{{\tau_{0}}^{2}} + u =0.
\end{gather*}
The general solution of the linear equation is
\begin{gather} \label{isog4}
 u(t)= \sqrt{A_0} \cos \left(B_0 - {\phi \over {\tau_0}} t\right),\qquad \phi
 =\arccos \left({{2-{\tau_0}^{2}}\over {2}}\right).
\end{gather}

The transformation \eqref{sal8} changes the f\/irst integrals
\eqref{int2} into the following ones:
\begin{gather*} 
{u_t}^2 + u u^{+} =A_0, \quad  \left( {\phi \over {\tau_0}} t
+{\phi\over {2}}\right)-  {1\over{2}} \arccos \left(\frac{2}{(1+
\frac{{u_t}^{2}}{u u^{+}})} - \cos \phi\right) =B_0.
\end{gather*}

 The solution family \eqref{sol22}, as well as the family \eqref{isog4}, are quite
 similar to those of the corresponding
dif\/ferential equations, and the superposition principle for two
solutions of the dif\/ference scheme \eqref{gridss2}, \eqref{eq15}
can be formulated in the same manner as for the continuous case.

Notice that the linearizing transformations presented above change
the admitted symmetries of the underlining equation. The way to
obtain such transformation could be viewed as follows. The f\/irst
step is construction of an invariant dif\/ference model (that was
done in \cite{1,2}) with the conservation of Lagrangian structure,
to make the Noether-type theorems applicable. Then, as the second
step, we f\/ind a dif\/ference non-point transformation, which
makes the nonlinear invariant mesh regular (since otherwise even
linear dif\/ference equation would be nonlinear). This step is
quite evident from the structure of the mesh equation. The third
step is to obtain a continuous limit of the transformation of
lattice, and then the corresponding tangent transformations in the
space $x$, $y$, $t$, $u$, which preserve both constants of the two
f\/irst integrals. As a~last step, we complete the appropriate
linearizing transformations for discrete models.

It should be pointed out that the  linearizing transformations are
not unique. Indeed, one can f\/ind another relation between
f\/irst integrals as far as any integral can be expressed by means
of any smooth function of two independent f\/irst integrals. We
should also add  that existence of two explicit f\/irst integrals
allows one to express the linearizing transformation explicitly
from the space $(x,y,y')$ to the space $(t,u,u')$. In Example 4 we
will deal with a generalization of equation \eqref{eq3.14}, which
has two-dimensional symmetry and the only f\/irst integral. In
that case we are able to construct linearizing transformations as
connections between dif\/ferentials.

In the following  examples we brief\/ly reproduce the linearizing
transformations for both invariant ODEs and corresponding
invariant dif\/ference equations and meshes.

\section{Example 2: ODE with 3-dimensional symmetry group}

The dif\/ferential equation
\begin{gather} \label{eq34c}
y'' =  {(y')} ^{ { k - 2 \over k-1} },\qquad k \neq 1
\end{gather}
has a symmetry algebra generated by the following operators
\begin{gather*}
X_{1}= {\ddx} , \qquad  X_{2}= {\ddy} , \qquad  X_{3}= x {\ddx} +
k y  {\ddy}, \qquad k \neq 0, \tfrac{1}{2},  1, 2 .
\end{gather*}
This equation can be obtained by the variational procedure from
the Lagrangian
\begin{gather*} 
L = { (k-1) ^{2} \over k} ( y' )^ {\textstyle { k \over k-1}}  + y
,
\end{gather*}
 which admits operators $X_{1}$ and $X_{2}$ for any
parameter $k \neq1$ , and therefore the Noether theorem yields two
f\/irst integrals
\[
 J_{1} = { (1-k) \over k} ( y' )^ {\textstyle { k
\over k-1}} + y = A^0, \qquad J_{2}  =  (k-1) ( y' )^{{ 1 \over
k-1} }  -x = B^0.
\]
Eliminating $y'$ we  f\/ind the general solution:
  \begin{gather*} 
y = {1 \over k} \left( { 1 \over k-1 } \right)^{(k-1)} ( x + B^{0}
) ^k  +  A^{0}.
\end{gather*}

 Following the same procedure as described in Example
1, one can f\/ind the non-point transformation
\begin{gather} \label{iso1+}
x= -V(t), \qquad y = t + {{k-1}\over{k}} (V')^{k} + (k-1)\ln V',
\qquad y'= (V')^{k-1},
\end{gather}
which transforms equation \eqref{eq34c} into the following linear
equation
\begin{gather} \label{iso2+}
(k-1) V'' +  V'=0.
\end{gather}
 The transformation \eqref{iso1+} changes the f\/irst integrals
into the following ones
\begin{gather*} 
(k-1)\ln V' + t=A_0, \qquad  (k-1) V' +  V =B_0.
\end{gather*}
The latter yields the general solution of the linear equation
\eqref{iso2+}:
\begin{gather*} 
 V(t)= (1-k) e^{{A_0 -t}\over{k-1}}  +B_0  .
\end{gather*}

As an invariant dif\/ference scheme for  the equation
\eqref{eq34c}, the following dif\/ference equations were
constructed \cite{2}:
\begin{gather*} 
{{2\alpha (k-1)}\over{h_{+}+ h_{-}}}
\left((y_{x})^{{1}\over{k-1}}-(y^{-}_{x})^{{1}\over{k-1}}\right)=1,
\qquad {{h_{+}}\over{(y_{x})^{{1}\over{k-1}}}}=
 {{h_{-}}\over{(y^{-}_{x})^{{1}\over{k-1}}}},
\end{gather*}
where the constant $\alpha$ should be chosen in accordance with
the compatibility condition for the f\/irst integrals (see
\cite{2} for details).

We again rewrite the transformation \eqref{iso1+} as a connection
between dif\/ferentials
\begin{gather*} 
dt = {{dx}\over{(y')^{{1}\over{k-1}}}}, \qquad
{{{(y')^{{1}\over{k-1}}}}= - V'},
\end{gather*}
 and then involve transformations as a two-point change of
 variables
\begin{gather*} 
{(y_{x})^{{1}\over{k-1}}}=- V_t, \qquad
{(y^{-}_{x})^{{1}\over{k-1}}}=- V^{-}_t, \qquad
\tau^{+}={{h_{+}}\over{(y_{x})^{{1}\over{k-1}}}}, \qquad \tau^{-}=
{{h_{-}}\over{(y^{-}_{x})^{{1}\over{k-1}}}}.
\end{gather*}

Evidently, the transformed mesh will be regular
\begin{gather*} 
\tau^{+}= \tau^{-}= \tau ^{0},
\end{gather*}
and the equation will be linear
\begin{gather*}
{2\alpha (k-1)}{{V_t - V^{-}_t}\over{\tau ^{0}}} + {{V_t +
V^{-}_t}\over{2}} =0.
\end{gather*}

\section{Example 3: ODE with a 3-dimensional symmetry group}

We consider one more three-dimensional group and its Lie algebra,
\begin{gather*} 
 \quad  X_{1}= {\ddx} , \qquad  X_{2}= {\ddy},
\qquad  X_{3}= x {\ddx} + (x+y ) {\ddy}.
\end{gather*}
The corresponding invariant second-order ODE is the following:
\begin{gather} \label{ww2}
y'' =  \exp ( -y' ).
\end{gather}
It can be obtained from the Lagrangian function
\begin{gather*} 
L = \exp( y' ) + y,
\end{gather*}
which is divergently invariant with respect to $X_{1}$, $X_{2}$.
The corresponding f\/irst integrals of the equation~(\ref{ww2})
are
\begin{gather} \label{ww5}
\exp ( y' )  - x  = A_0 , \qquad \exp ( y' ) ( 1 - y' ) + y =
 B_0.
\end{gather}
So, the general solution of equation~(\ref{ww2}) is
\begin{gather*}    
y = (x +  B_0) ( \ln ( x+ B_0 ) - 1 ) + A_0.
\end{gather*}

The equation \eqref{ww2} can be linearized by the non-point
transformation
\begin{gather} \label{so1}
x= V(t), \qquad y = (1+V')\ln V'- V' -t, \qquad y'= \ln V',
\end{gather}
which transforms the equation \eqref{ww2} into the following
linear one
\begin{gather*} 
V''- V' =0.
\end{gather*}
 The transformation \eqref{so1} changes the f\/irst integrals \eqref{ww5}
into
\begin{gather*} 
V'- V =A_0, \qquad \ln V'- t =B_0,
\end{gather*}
which yield the general solution of the linear equation,
\begin{gather*} 
 V(t)= e^{B_0 + t}- A_0.
\end{gather*}
 The superposition principle for solutions of \eqref{ww2} can be
established similarly to Example~1.

The invariant dif\/ference model  constructed in \cite{2} is the
following:
\begin{gather}\label{eq47}
  \frac{\beta}{h_{+} }(e^{y_x} - e^{y^{-}_x})=1,
\\
{{h_{+}}\over{e^{y_x}}} = {{h_{-}}\over{e^{y^{-}_x}}},\label{eq48}
\end{gather}
where
\[
\beta=
\frac{e^{1+\varepsilon^{2}}}{(1+\varepsilon)^{1+\frac{1}{\varepsilon}}},
\qquad {{h_{+}}\over{e^{y_x}}}= \varepsilon = {\rm const} >0,
\]
the constants $\beta$, $\varepsilon$ are chosen in accordance with
the compatibility condition for the f\/irst integrals.

 The above dif\/ference model preserves both all symmetries of
the original dif\/ferential equation and the Lagrangian structure
(see \cite{2}), and therefore can be integrated. The general
solution is
\[
y = (x +  B_0)  \ln ( x+ B_0 ) - (1+\varepsilon^{2})(x +  B_0)  +
A_0, \qquad  h_+= \varepsilon(x+B_0).
\]

 This dif\/ference model can also be linearized. We rewrite the transformation \eqref{so1} as
\begin{gather*} 
 V'= e^{y'}, \qquad dt = {{dx}\over{e^{y'}}},
\end{gather*}
and f\/ind the corresponding dif\/ference transformations as
\begin{gather*}
V_t = {e^{y_x}}, \qquad V^{-}_t = {e^{y^{-}_x}}, \qquad \tau^{+}=
{{h^{+}}\over{e^{y_x}}}, \qquad
\tau^{-}={{h^{-}}\over{e^{y^{-}_x}}}.
\end{gather*}
 Applying the above transformation we linearize the
dif\/ference equation \eqref{eq47} together with the mesh
\eqref{eq48},
\begin{gather*} 
 {{V^+ -2V+ V^-}\over{{\tau_0}^2}} - {{V^{+} -  V^-}\over{{2\tau_0}}}= 0,
\qquad 
 \tau^{+} = \tau^{-}= \tau_0.
\end{gather*}

The last fourth Lie's equation, which possesses 3-dimensional
symmetry, contains an arbitrary constant, and it is a f\/irst
integral of some nonlinear third-order ODE (the dif\/ferential
consequence of this second-order ODE). The linearization of this
equation is rather complicated and is not considered here.

\section{Example 4: ODE with a 2-dimensional symmetry group}

S.~Lie identified two classes of ODEs which possess a
2-dimensional symmetry group. Both classes contain arbitrary
functions. For example, one class of equations is the following
\[
y'' = \frac{1}{x} F(y'),
\]
which is invariant with respect to the group with the following
operators
\[
X_{1}= {\dx y} , \qquad  X_{2}=  x {\dx x} + y {\dx y}.
\]

It seems to be impossible to linearize the whole class of
invariant ODEs, thus we consider just representative equations. We
will choose the example which is important from  the physical
point of view.

 First, we interchange dependent and independent variables and,
 second, we make scaling of the independent variable to obtain
 another representation of the underlining class of equations:
\[
y'' = y^{n}F(x^2 y^{n-1}),
\]
where $n$ is some constant, $n \neq -1$.

Now we consider the special case of the above equation, when the
function $F=1$. It will be the general case of the power
nonlinearity, i.e. the  dif\/ferential equation
\begin{gather} \label{equ}
y'' =  y^{n}, \qquad n \neq -1,
\end{gather}
 which possesses an isomorphic two-dimensional Lie
algebra of operators
\begin{gather} \label{o1}
X_{1}= {\dx x} , \qquad  X_{2}= (n-1) x {\dx x} -2 y {\dx y}.
\end{gather}

  Notice, that this example makes
sense from the physical point of view, as equation (\ref{equ}) can
be considered as a dynamical equation for a particle in a f\/ield
of the power potential. This equation can be obtained from the
Lagrangian function
\[
L = {y'}^2   + {{2}\over{n+1}}{ y ^{n+1} },
\]
which admits the only operator $X_{1}$.

In contrast to  Example 1, which is a special case of equation
\eqref{equ} with $n=-3$, the underlining equation for an arbitrary
$n$ has a unique f\/irst integral:
\begin{gather*} 
 J_{1} =  {y'}^2   - {{2}\over{n+1}}{ y ^{n+1} } =A^0.
\end{gather*}

The second  integral can be expressed by a quadrature
\[
x - \int \frac{dy}{\sqrt{A_0 + \frac{2}{n+1} y^{n+1}}} = B_0,
\]
and for the only cases
\[
n = \frac{1}{k} - 1, \qquad  n = \frac{1-2k}{1+2k},
\]
where $k$ is integer, this integral can be expressed by a f\/inite
superposition of  elementary functions. We see that equation
\eqref{eq3.14}, considered in Example~1, corresponds to $k=-1$. In
contrast to a~general case, equation \eqref{eq3.14} is the only
equation of power nonlinearity which possesses a 3-di\-mensional
symmetry group. The exceptional potential which corresponds to
equation \eqref{eq3.14} is known in theoretical physics.

Thus, the example of the general power potential under
consideration is suf\/f\/iciently dif\/ferent from the above
three, as we have fewer symmetries and the only f\/irst explicit
integral. Never\-theless, equation (\ref{equ}) for any $n$ can be
linearized by a non-point transformation. The only disadvantage is
that we will express the linearizing transformation in terms of
dif\/ferentials.

Again, we would like  to preserve the f\/irst integral under
linearization, for example, as follows:
\[
 {y'}^2   - {{2}\over{n+1}}{ y ^{n+1} } =A^0 = {u'}^2 - u^2 .
\]
Then, we can choose the change of dependent variables and
derivatives
\[
{{2}\over{n+1}}{ y ^{n+1} }= u^2, \qquad y'(x) = u'(t),
\]
and, therefore, the transformation of dif\/ferentials and
derivatives could be the following
\begin{gather} \label{isok1}
dx = |\frac{n+1}{2}|^{-{n\over{n+1}}} {u}^{{1-n}\over{n+1}} dt,
\qquad y(x) = |\frac{n+1}{2}|^{1\over{n+1}} {u}^{2\over{n+1}},
\qquad y'(x) = u'(t).
\end{gather}
 The transformation (\ref{isok1}) transforms the equation
(\ref{equ}) into the following linear one
\begin{gather} \label{isoj2}
u''- u =0,
\end{gather}
 which has the f\/irst integral
\begin{gather*}
{u'}^2 - u^2 =A_0.
\end{gather*}
 Meanwhile the equation (\ref{isoj2}) has one more
functionally independent integral
\begin{gather*}
({u'} +u)e^{-t} =B_0,
\end{gather*}
 and therefore the general solution is
\begin{gather*} 
 u(t)= A e^{t} + B e^{-t}.
\end{gather*}
Notice that one can transform backwards the above general solution
and the second integral just for exceptional values of~$n$.

We will construct an invariant dif\/ference model by means of full
set of dif\/ference invariants of the Lie algebra (\ref{o1}) on a
dif\/ference stencil:
\begin{gather} \label{gru1}
I_1= {h_{+}\over{h_-}},\qquad I_2= {{y_{+}}\over{y}},\qquad I_3=
{{y_{-}}\over{y}}, \qquad I_4= h_{+}
{y_{+}}^{{n-1\over{4}}}{y}^{{n-1\over{4}}} .
\end{gather}

 As an invariant dif\/ference mesh, we will use the following relation
between dif\/ference invariants:
\begin{gather} \label{grido}
 h_{+}  {y_{+}}^{{n-1\over{4}}}=h_{-}  {y_{-}}^{{n-1\over{4}}},
\end{gather}
  which has the integral
\begin{gather*} 
 h_{+}  {y_{+}}^{{n-1\over{4}}}{y}^{{n-1\over{4}}}  = \varepsilon,
 \qquad \varepsilon = {\rm const}.
\end{gather*}

By means of dif\/ference invariants \eqref{gru1} one can develop
an invariant approximation of equation~\eqref{equ} on mesh
(\ref{grido}):
\begin{gather}\label{eq64}
\left(\frac{{y_{+}}^{{n+1\over{2}}} -{y_{}}^{{n+1\over{2}}}
}{h_{+}{y_{+}}^{{n-1\over{4}}}} -\frac{{y_{}}^{{n+1\over{2}}}
-{y_{-}}^{{n+1\over{2}}} }{h_{-}{y_{-}}^{{n-1\over{4}}}}\right)
\frac {1}{h_{+}}={y_{+}}^{{n-1\over{4}}}{y}^{n}.
\end{gather}

As the dif\/ference analog of the transformation \eqref{isok1}, we
will use the non-point transformations for discrete variables as
follows
\begin{gather*} 
u(t)=|{{n+1}\over{2}}|^{-{1\over{2}}}{y_{}}^{{n+1}\over{2}},
\qquad
u^{+}=|{{n+1}\over{2}}|^{-{1\over{2}}}{y_{+}}^{{n+1}\over{2}},
\qquad
u^{-}=|{{n+1}\over{2}}|^{-{1\over{2}}}{y_{-}}^{{n+1}\over{2}},\\
\tau^{+}=h^{+}{y_{}}^{{n-1}\over{4}} {y_{+}}^{{n-1}\over{4}},
\qquad \tau^{-}=h^{-}{y_{}}^{{n-1}\over{4}}
{y_{-}}^{{n-1}\over{4}},\nonumber
\end{gather*}
where $\tau^{+}$ and $\tau^{-} $ are the new mesh steps.

Invariant mesh \eqref{grido} is transformed into the regular one
\begin{gather*} 
\tau^{+}=\tau^{-}=\tau_{0}=\varepsilon,
\end{gather*}
and the nonlinear equation \eqref{eq64} becomes linear,
\begin{gather} \label{sal4}
\frac{u^+ -2u + u^-}{{\tau_{0}}^{2}} - u =0.
\end{gather}

Equation \eqref{sal4} has the following dif\/ference integrals
\begin{gather*}
{u_t}^2 - u u^{+} =A_0, \qquad (u_t  + \beta
u^+)q^{{t\over{\tau_0}}} =B_0,
\end{gather*}
where
\[
\beta = \frac{\tau_0 + \sqrt{{\tau_0}^2  +4}}{2}, \qquad q=\frac{-
\tau_0 + \sqrt{{\tau_0}^2  +4}}{\tau_0 + \sqrt{{\tau_0}^2  +4}}.
\]

The general solution of \eqref{sal4}  can be found from the above
integrals. Alternatively, one can look for a solution in the form
\begin{gather}\label{eq69}
u(t) = A {q_{1}}^{t\over{\tau_0}} + B {q_{2}}^{{t\over{\tau_0}}} .
\end{gather}
Substituting \eqref{eq69} into \eqref{sal4} yields the needed
constants
\begin{gather*}
{q_{1,2}}= \frac{{\tau_0}^2 +2 \pm {\tau_0}\sqrt{{\tau_0}^2
+4}}{2}.
\end{gather*}

\section{Example 5: special case of an ODE\\ with a 2-dimensional
symmetry group}

 For the special case $n= -1$ the corresponding ODE
\begin{gather} \label{23}
y'' = y^{-1},
\end{gather}
 possesses  the two-dimensional  Lie algebra given by the operators
\begin{gather*} 
X_{1}= {\dx x} , \qquad  X_{2}=  x {\dx x} + y {\dx y}.
\end{gather*}
 Equation~(\ref{23}) can be obtained as the Euler equation from the
  Lagrangian
\begin{gather*} 
L = \ln y + \frac{1}{2} (y')^2
\end{gather*}
which is invariant with respect
 to $X_{1}$.

The corresponding f\/irst integral of equation~(\ref{23}) is
\begin{gather} \label{wa5}
\ln y - \frac{1}{2} (y')^2 = A_0 ,
\end{gather}
 while the second integral could be expressed by a
quadrature
\[
x - \int \frac{dy}{\sqrt{A_0 + 2 \ln y}} = B_0.
\]

Following the way of preservation of the f\/irst integral under
the linearizing transformations
\begin{gather} \label{waz5}
\ln y - \frac{1}{2} (y')^2 = A_0 = u - \frac{1}{2}(u')^2,
\end{gather}
 we choose the change of dependent variable and derivative as follows
\begin{gather*} 
\ln y(x)=  u(t), \quad  \quad y'(x) =u'(t).
\end{gather*}

Then the independent variables are connected as
\[
dx = e^{u}dt.
\]

It follows from f\/irst integral~(\ref{waz5}) that the
second-order ODE is the following:
\begin{gather} \label{solo2}
u''=1.
\end{gather}

Thus, equation~(\ref{23}) can be linearized by a non-point
transformation
\begin{gather} \label{solo1}
y(x)= e^{ u(t)}, \qquad  dx = e^{u}dt, \qquad y'=u',
\end{gather}
which transforms equation~(\ref{23}) into the  linear
ODE~(\ref{solo2}).

The transformation~(\ref{solo1}) changes the f\/irst integral
\eqref{wa5} into the following:
\begin{gather*} 
u - \frac{1}{2}(u')^2  =A_0.
\end{gather*}
For the linear equation~(\ref{solo2}) there exists one more
independent f\/irst integral
\begin{gather*} 
     u'- t =B_0,
\end{gather*}
 which yields the general solution of the linear equation in the
 form
\begin{gather} \label{solo5}
 u(t)= \frac{(t+B_0)^2}{2} + A_0.
\end{gather}

Thus,  the linearizing transformation~(\ref{solo1}) yields an
implicit superposition principle for equation~(\ref{23}).

An invariant dif\/ference model can be obtained from  the
following dif\/ference Lagrangian:
\begin{gather*} 
L = \ln y +\ln{y^{+}} +  (y_x)^2,
\end{gather*}
where $y_x= \frac{y^{+} -y}{h_{+}} $ is a right dif\/ference
derivative.

The application of the dif\/ference analog of the Noether
theorem~\cite{5} yields the appropriate quasi-extremal equation
\begin{gather}\label{ex1}
y_x - {y_x}^-= \frac{\ln y^{+} -\ln y^{-}}{y_x + {y_x}^-},
\end{gather}
where ${y_x}^{-}= \frac{y- y^{-}}{h_{-}} $ is a left dif\/ference
derivative, and the f\/irst integral
\begin{gather*}
\ln ({y^{+} y}) - {y_x}^2= 2 A.
\end{gather*}

The invariant dif\/ference mesh for~(\ref{ex1}) should be
expressed by means of dif\/ference invariants
\begin{gather*}
h^{+} =
h^{-}F\left(\frac{y^{+}}{y},\frac{y}{y^{-}},\frac{h^{-}}{y}\right),
\end{gather*}
and it is not yet def\/ined.

As a dif\/ference model for~(\ref{solo2}) we consider the
following linear second-order equation
\begin{gather} \label{volk1}
\frac{u^{+} -2u +u^{-}}{\tau^{+}} =\frac{1}{\tau^{+}}(u_t -
{u_t}^{-}) = 1,
\end{gather}
 on regular lattice
\begin{gather} \label{ok1}
{\tau^{+}} ={\tau^{-}}.
\end{gather}
 The last model has the following f\/irst integrals:
\begin{gather} \label{volk8}
 (u^{+} + u) - {u_t}^2= 2 A,
\\ \label{volk9}
  {u_t} - \left(t+ \frac{\tau^{+}}{2}\right)= B,
\end{gather}
and mesh equation~(\ref{ok1}) has an evident integral
\[
{\tau^{+}} ={\tau^{-}}=\tau_0= {\rm const}.
\]

From two f\/irst integrals (\ref{volk8}), (\ref{volk9}) one can
exclude $u^{+}$ and obtain a general solution
\begin{gather*} 
 u(t)= \frac{(t+B)^2}{2} - \frac{{\tau_{0}}^{2}}{8}   + A,
\end{gather*}
which well corresponds to the solution~(\ref{solo5}) of
ODE~(\ref{solo2}).

Now we can write down the linearizing transformation
\begin{gather} \label{vov1}
 u= \ln y, \qquad u^+= \ln y^+, \qquad u^-= \ln y^-, \qquad u_t=y_x, \qquad
 {u_t}^-={y_x}^-,
 \end{gather}
\[
 \tau^+=\frac {\ln y^{+} - \ln y }{y^+ - y} h^{+}, \qquad
 \tau^+=\frac {\ln y - \ln y^{-} }{y - y^-} h^{-},
\]
which preserves both the f\/irst integral~(\ref{volk8}) of
equation~(\ref{volk1}) and the integral of a mesh. Notice that in
the continuous limit transformation~(\ref{vov1}) tends
to~(\ref{solo1}). Under the action of~(\ref{vov1}) the regular
mesh~(\ref{ok1}) is transformed into the following non-regular
invariant mesh
\begin{gather} \label{volg1}
h^{+} \frac{\ln{\frac{y^{+}}{y}}}{y^{+}-y} = h^{-}
\frac{\ln{\frac{y}{y^{-}}}}{y-y^{-}},
 \end{gather}
which completes the dif\/ference model for ODE~(\ref{23}).

Thus, invariant model~(\ref{ex1}), (\ref{volg1}) is connected with
the linear model~(\ref{volk1}), (\ref{ok1}) by two-points
transformation~(\ref{vov1}).

The remaining class of  second-order ODEs from the Lie's list
contains all autonomous equations which are invariant with respect
to the only operator $X_{1}= {\dx x}. $ For this class the
well-known procedure reduces the order of equation by one. A
similar reduction can be done for any autonomous dif\/ference
equation on a lattice which is invariant with respect to a
translation of~$x$.

\section{Concluding remarks}

It was shown that, for all  examples of  invariant dif\/ferential
equations from the Lie's list,  there exist non-point linearizing
transformations (which transform the family of solutions of the
original nonlinear ODE into the family of solutions of a linear
one) and, therefore, nonlinear superposition principle for their
solutions.

The procedure that we followed in this article is based on the
preservation of at least one f\/irst integral for both the ODE and
the dif\/ference equation, and a transformation of the invariant
nonlinear dif\/ference mesh into a regular one. We started from
the continuous case, where second-order ODEs possessing symmetry
were already known. From \cite{1,2} we know the symmet\-ry
preserving the dif\/ference model, i.e., the dif\/ference equation
and a mesh, which in addition conserves the Lagrangian structure
of the original ODE. Then we choose the change of dependent
variables which preserves the known f\/irst integral. These
transformations can be dif\/ferent for dif\/ferent representations
of f\/irst integrals, and we choose such change of variables,
which makes an invariant mesh regular. Then we obtain a continuous
limit of these transformations of a~lattice and derive the
corresponding transformations in the space $(y,dx,u,dt)$,
preserving all f\/irst integrals. For the cases of two explicit
integrals (Examples~1,~2,~3), we express explicitly the change of
variables in the space $(x,y,y',t,u,u')$. Finally, we complete the
appropriate lineari\-zing transformations for the discrete model.
Notice that the developed non-point linearizing transformations
are not contact transformation in the space $(x,y,y',t,u,u')$, but
are such on the family of solutions of invariant ODEs.

It worth mentioning that all dif\/ference and dif\/ferential
equations considered here are integrable (at least in quadratures)
as far as they possess a symmetry algebra of the dimension not
smaller than two. Thus, the integrability is not of interest for
these equations. The result is the explicit non-point
linearization of integrable nonlinear dif\/ferential and
dif\/ference models.  The second valuable result of the
linearization is a nonlinear superposition principle for solutions
of nonlinear equations, which have def\/inite physical meaning.
Finally, we can indicate that all mentioned properties can be
preserved in f\/inite dif\/ference models.

\subsection*{Acknowledgments}

The author thanks P.~Winternitz and E.~Ferapontov for helpful
discussions and remarks. The author's research was sponsored in
part by the Russian Fund for Basic Research under the research
project No~06-01-00707.

\LastPageEnding

\end{document}